\documentclass[review,12pt]{elsarticle}

\usepackage{amssymb}
\usepackage{amsthm}
\usepackage{amsmath,color}
\usepackage{mathrsfs}
\usepackage{graphicx}
\usepackage{epstopdf}
\usepackage{float}
\usepackage{caption}
\usepackage{subcaption}
\usepackage{bm}
\usepackage{bbm}
\usepackage{mathrsfs}
\usepackage{cleveref}
\usepackage{soul}
\usepackage{accents}
\usepackage{color,soul} 
\usepackage{color} 
\biboptions{sort&compress}
\newsavebox{\measurebox} 

\journal{Theoretical and Applied Fracture Mechanics}

\makeatletter
\def\@author#1{\g@addto@macro\elsauthors{\normalsize%
    \def\baselinestretch{1}%
    \upshape\authorsep#1\unskip\textsuperscript{%
      \ifx\@fnmark\@empty\else\unskip\sep\@fnmark\let\sep=,\fi
      \ifx\@corref\@empty\else\unskip\sep\@corref\let\sep=,\fi
      }%
    \def\authorsep{\unskip,\space}%
    \global\let\@fnmark\@empty
    \global\let\@corref\@empty  
    \global\let\sep\@empty}%
    \@eadauthor={#1}
}
\makeatother

\begin{document}

\begin{frontmatter}



\title{Damage modeling in Small Punch Test specimens}


\author{E. Mart\'{\i}nez-Pa\~neda \corref{cor1}\fnref{DTU}}
\ead{mail@empaneda.com}

\author{I.I. Cuesta\fnref{UBU}}

\author{I. Pe\~nuelas\fnref{Uniovi}}

\author{A. D\'{\i}az\fnref{UBU}}

\author{J.M. Alegre\fnref{UBU}}

\address[DTU]{Department of Mechanical Engineering, Solid Mechanics, Technical University of Denmark, DK-2800 Kgs. Lyngby, Denmark}

\address[UBU]{Structural Integrity Group, University of Burgos, Escuela Polit\'{e}cnica Superior. Avenida Cantabria s/n, 09006, Burgos, Spain}

\address[Uniovi]{Department of Construction and Manufacturing Engineering, University of Oviedo, Gij\'on 33203, Spain}

\cortext[cor1]{Corresponding author.}

\begin{abstract}
Ductile damage modeling within the Small Punch Test (SPT) is extensively investigated. The capabilities of the SPT to reliably estimate fracture and damage properties are thoroughly discussed and emphasis is placed on the use of notched specimens. First, different notch profiles are analyzed and constraint conditions quantified. The role of the notch shape is comprehensively examined from both triaxiality and notch fabrication perspectives. Afterwards, a methodology is presented to extract the micromechanical-based ductile damage parameters from the load-displacement curve of notched SPT samples. Furthermore, Gurson-Tvergaard-Needleman model predictions from a top-down approach are employed to gain insight into the mechanisms governing crack initiation and subsequent propagation in small punch experiments. An accurate assessment of micromechanical toughness parameters from the SPT is of tremendous relevance when little material is available.\\
\end{abstract}


\end{frontmatter}


\section{Introduction}
\label{Introduction}

Many engineering applications require a mechanical characterization of industrial components from a limited amount of material. Under such circumstances, it is often not possible to obtain specimens of the dimensions demanded by standard testing methodologies. With the aim of overcoming this hurdle, a miniature non-standard experimental device was developed in the early 80s \cite{M81}. The aforementioned testing methodology, commonly known as Small Punch Test (SPT), employs very small specimens (generally, 8 mm diameter and 0.5 mm thickness) and may be considered as a non-destructive experiment. The SPT has consistently proven to be a reliable tool for estimating the mechanical \cite{R09,G14} and creep \cite{DM09,A16} properties of metallic materials and its promising capabilities in fracture and damage characterization have attracted great interest in recent years (see, e.g., \cite{HB92,J03,AK06,C08,L08,P09,C10,C11,R13,X14,S16,M16}).\\

\begin{figure}[H]
\centering
\includegraphics[scale=0.45]{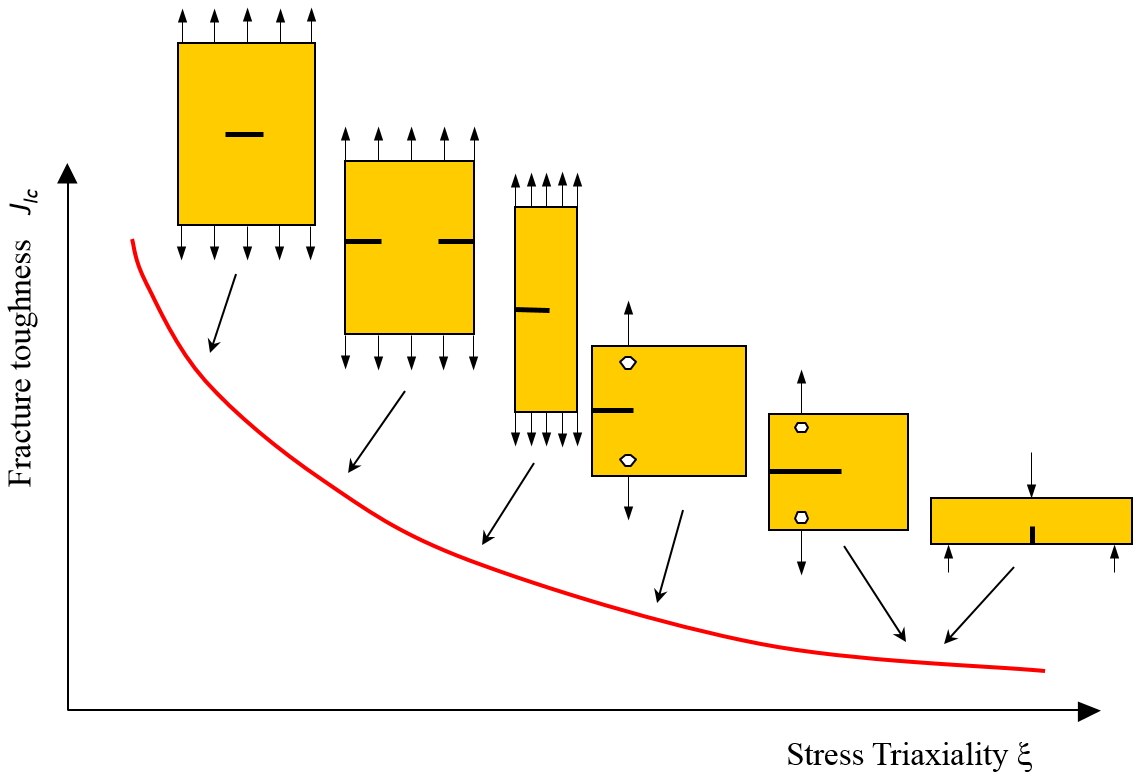}
\caption{Influence of the specimen configuration on fracture toughness.}
\label{fig:Triaxiality}
\end{figure}

Although brittle fracture has been observed in certain materials at low temperatures \cite{L08,X14,S16}, the stress state inherent to the SPT favors ductile damage. It therefore comes as no surprise that efforts to characterize the initiation and subsequent propagation of cracks in SPT specimens have mostly employed models that account for the nucleation, growth and coalescence of microvoids (see, e.g., \cite{AK06,C08,P09,C10,M16} and references therein). The model by Gurson \cite{G75}, later extended by Tvergaard and Needleman \cite{TN84}, is by far the most frequent choice, but other models - such as the one by Rousselier \cite{R87} - have also been employed \cite{C08}. These  models are able to quantitatively capture the experimental results by fitting several parameters that account for the ductile damage mechanisms taking place. A variety of inverse techniques - including the use of evolutionary genetic algorithms \cite{P09,C10,P11} and neural networks \cite{AK06} - have been proposed to compute the Gurson-Tvergaard-Needleman (GTN) \cite{G75,TN84} parameters from the load-displacement curve of unnotched SPT specimens. Void-based models have been particularly helpful in the development of new methodologies to estimate fracture toughness from SPT specimens \cite{M16}. However, some relevant aspects remain to be addressed. The substantially different constraint conditions attained in the SPT, relative to conventional testing procedures, constitute the most important problem to overcome. As depicted in Fig. \ref{fig:Triaxiality}, the high triaxiality levels (defined as the ratio of the hydrostatic stress to the von Mises equivalent stress) of standardized fracture toughness experiments - such as compact tension or three point bending tests - translate into conservative estimations of the fracture resistance. This is not the case of the SPT, hindering a direct comparison and leading to predictions that may significantly differ from the plane strain fracture toughness. Hence, current research efforts are mainly devoted to the development of notched or cracked SPT samples with the aim of increasing the attained triaxiality level \cite{J03,M16}. 

In this work, the influence of the shape of the notch on the SPT response is extensively investigated, considering both the constraint conditions and the fabrication process. Crack initiation and subsequent propagation is computed by means of the GTN model for various geometries of notched SPT specimens and results are compared to experimental data. Different methodologies to extract the micromechanical-based ductile damage parameters are proposed and the past, present and future capabilities of the SPT to characterize fracture and damage are thoroughly discussed.

\section{Experimental methodology}
\label{SPT description}

\begin{figure}[H]
\centering
\includegraphics[scale=0.7]{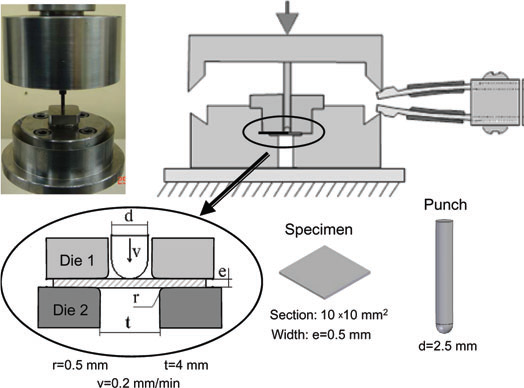}
\caption{Device and schematic description of the Small Punch Test}
\label{fig:SPT}
\end{figure}

The SPT employs a miniature specimen whose entire contour is firmly pressed between two dies with the load being applied at the center by means of a 2.5 mm hemispherical diameter punch. The special device outlined in Fig. \ref{fig:SPT} is coupled to a universal testing machine. A free-standing
extensometer is attached to the experimental device to accurately measure the punch displacement. The experiments are performed at room temperature with a punch speed of $v=0.2$ mm/min. Lubrication is employed to minimize the effects of friction.

\begin{figure}[H]
\centering
\includegraphics[scale=0.4]{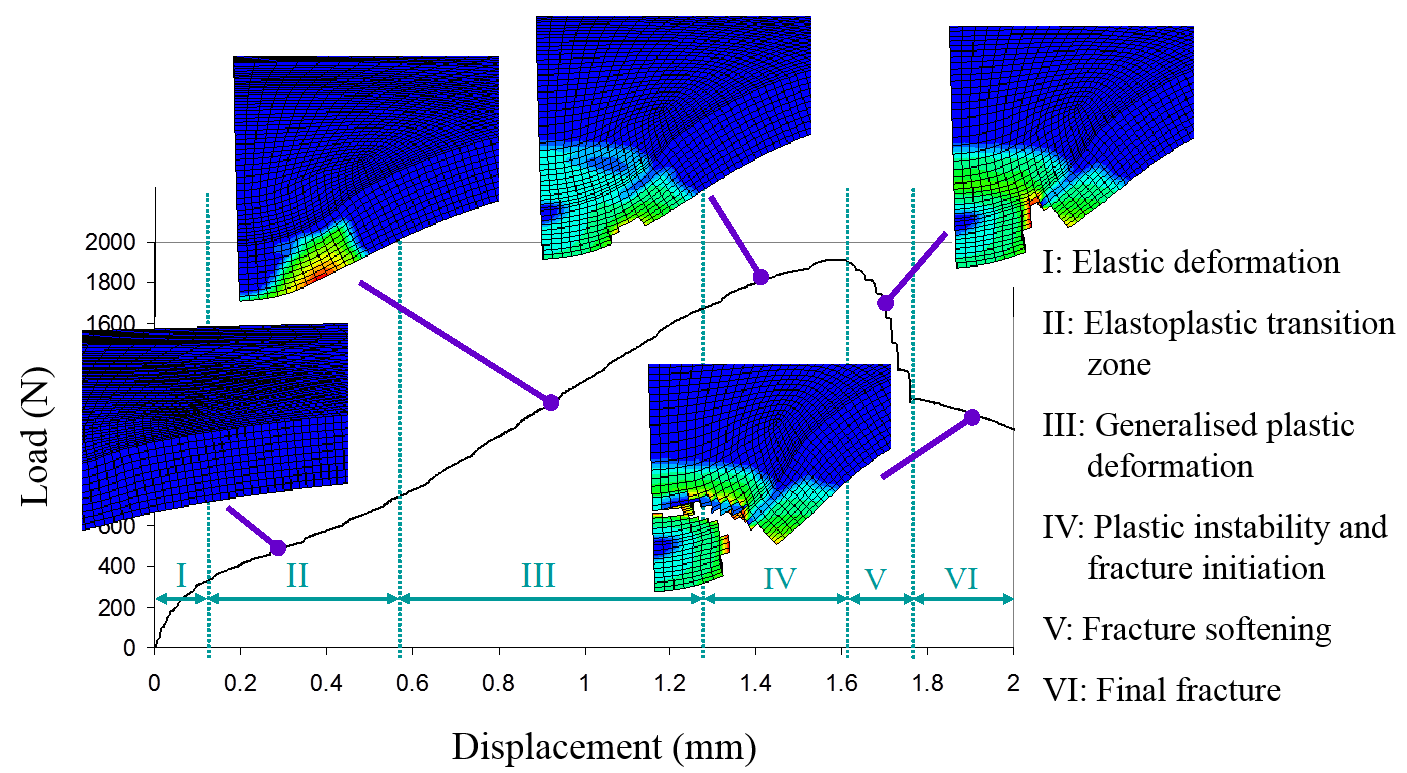}
\caption{Regions of the load - punch displacement curve in a Small Punch Test}
\label{fig:LoadvsDispTypical}
\end{figure}

The mechanical response of the SPT specimen is therefore characterized by means of the measured applied load versus punch displacement curve. Fig. \ref{fig:LoadvsDispTypical} shows the different stages that can be identified in the characteristic SPT curve of a material behaving in a ductile manner. Different criteria have been proposed to estimate mechanical and damage material parameters from the curve \cite{R09,R13}.

\section{Gurson-Tvergaard-Needleman model}
\label{GTN}

The influence of nucleation, growth and coalescence of microvoids is modeled by means of the well-known Gurson-Tvergaard-Needleman (GTN) \cite{G75,TN84} ductile damage model. Within the aforementioned framework, the yield function is defined by,
\begin{equation}
\Phi \left( \sigma_e, \sigma_h , \sigma_y , f \right)= \left( \frac{\sigma_e}{\sigma_y} \right)^2 + 2 q_1 f^* \textnormal{cosh} \left( \frac{3 q_2 \sigma_h}{2 \sigma_y} \right) - \left(1+ q_3 {f^*}^2\right) =0
\end{equation}

\noindent where $f$ is the microvoid volume fraction, $\sigma_h$ is the hydrostatic stress, $\sigma_e$ is the conventional Von Mises equivalent stress, $\sigma_y$ is the yield stress of the matrix material and $q_1$, $q_2$ and $q_3$ are fitting parameters as defined by Tvergaard \cite{T82}. The modified void volume fraction $f^*$ was introduced by Tvergaard and Needleman \cite{TN84} to model the decrease in load carrying capacity that accompanies void coalescence, such that,
\begin{equation} f^* =
  \begin{cases}
    f & \quad \text{for } f \le f_c \\
    f_c+\frac{f_u^*-f_c}{f_f-f_c} \left(f-f_c \right) & \quad \text{for } f > f_c\\
  \end{cases}
\end{equation}

\noindent with $f_c$ being the critical void volume fraction, $f_f$ the void volume fraction at final fracture and $f^*_u=1/q_1$ the ultimate void volume fraction. The current void volume fraction $\dot{f}$ evolves as a function of the growth rate of existing microvoids and the nucleation rate of new microvoids
\begin{equation}
\dot{f}=\dot{f}_{growth}+\dot{f}_{nucleation}
\end{equation}

\noindent where, according to Chu and Needleman \cite{CN80}, the latter is assumed to follow a Gaussian distribution, given by,
\begin{equation}
\dot{f}_{nucleation}=A \dot{\bar{\varepsilon}}^p
\end{equation}

\noindent with $\dot{\bar{\varepsilon}}^p$ being the equivalent plastic strain rate, and,
\begin{equation}
A= \frac{f_n}{S_n \sqrt{2 \pi}} \textnormal{exp} \left(-\frac{1}{2} \left(\frac{\bar{\varepsilon}^p-\varepsilon_n}{S_n} \right)^2 \right)
\end{equation}

\noindent Here, $\varepsilon_n$ is the mean strain, $S_n$ is the standard deviation and $f_n$ is the void volume fraction of nucleating particles.\\

Different methodologies have been proposed to fit model parameters from a variety of experimental tests (see, e.g., \cite{AK06,P09,M16}). A common procedure in the literature is to assume constant values of the parameters $q_1$ and $q_2$ (with $q_3=q_1^2$) based on the micromechanical cell studies by Tvergaard \cite{T81,T82}, but more complex models have also been proposed \cite{VF09}.

\section{Results}
\label{FE Results}

A numerical model of the SPT is developed by means of the finite element software Abaqus/Standard. Attending to the specimen geometry and test setup, quasi-static conditions are assumed and a 3-D approach is adopted, taking advantage of symmetry when possible. As described elsewhere \cite{G15b,M16}, 8-node linear brick elements are employed, with the mesh gradually being refined towards the notch, where the characteristic element length is determined from a sensitivity study. The lower matrix, the fixer and the punch are modeled as rigid bodies and their degrees of freedom are restricted except for the vertical displacement of the punch. The friction coefficient was set to $\mu=$0.1, which is a common value for steel-to-steel contact under partial lubrication. Ductile damage is captured by means of the GTN model, which is implemented in ABAQUS by means of a UMAT subroutine, where the consistent tangent moduli is computed through the Euler backward algorithm, as proposed by Zhang \cite{Z95}.\\

As discussed before, focus is placed in notched SPT specimens, as introducing a defect in the sample paves the way to establishing a direct correlation with standardized tests and allows for fracture resistance predictions applicable to a wide range of stress states. Hence, different geometries are modeled as a function of the various types of notches considered. 

\subsection{The role of the notch geometry}
\label{SubSec:TriaxialityStudy}

\begin{figure}[H]
\centering
\includegraphics[scale=0.1]{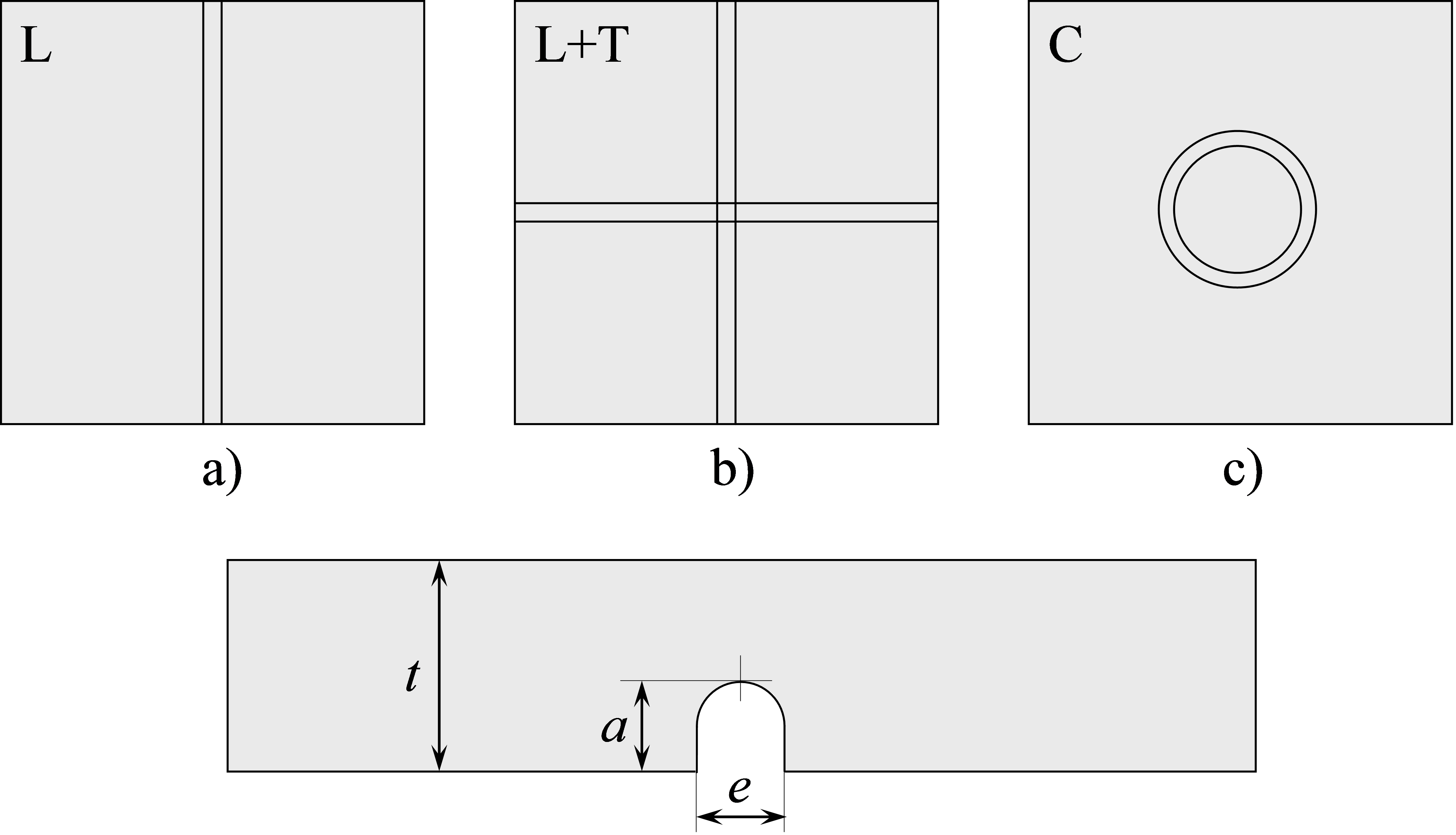}
\caption{Different notched SPT specimens examined. In all cases the notch radius equals $e/2=$100 $\mu m$.}
\label{fig:SPTconfigurations}
\end{figure}

The influence of the notch geometry on the stress triaxiality is thoroughly examined. Thus, as depicted in Fig. \ref{fig:SPTconfigurations}, three different notch classes have been considered; $10$x$10$ mm$^2$ square specimens with (i) a longitudinal notch (L), (ii) a longitudinal and transverse notch (L+T), and (iii) a circular notch of 3 mm diameter (C). Furthermore, for each geometry calculations are performed for two thicknesses ($t=0.5$ mm and $t=1$ mm) and four notch depths ($a/t=0.2$, $a/t=0.3$, $a/t=0.4$ and $a/t=0.5$). Hence, a total of 24 different configurations have been examined.

\begin{figure}[H]
\centering
\includegraphics[scale=0.8]{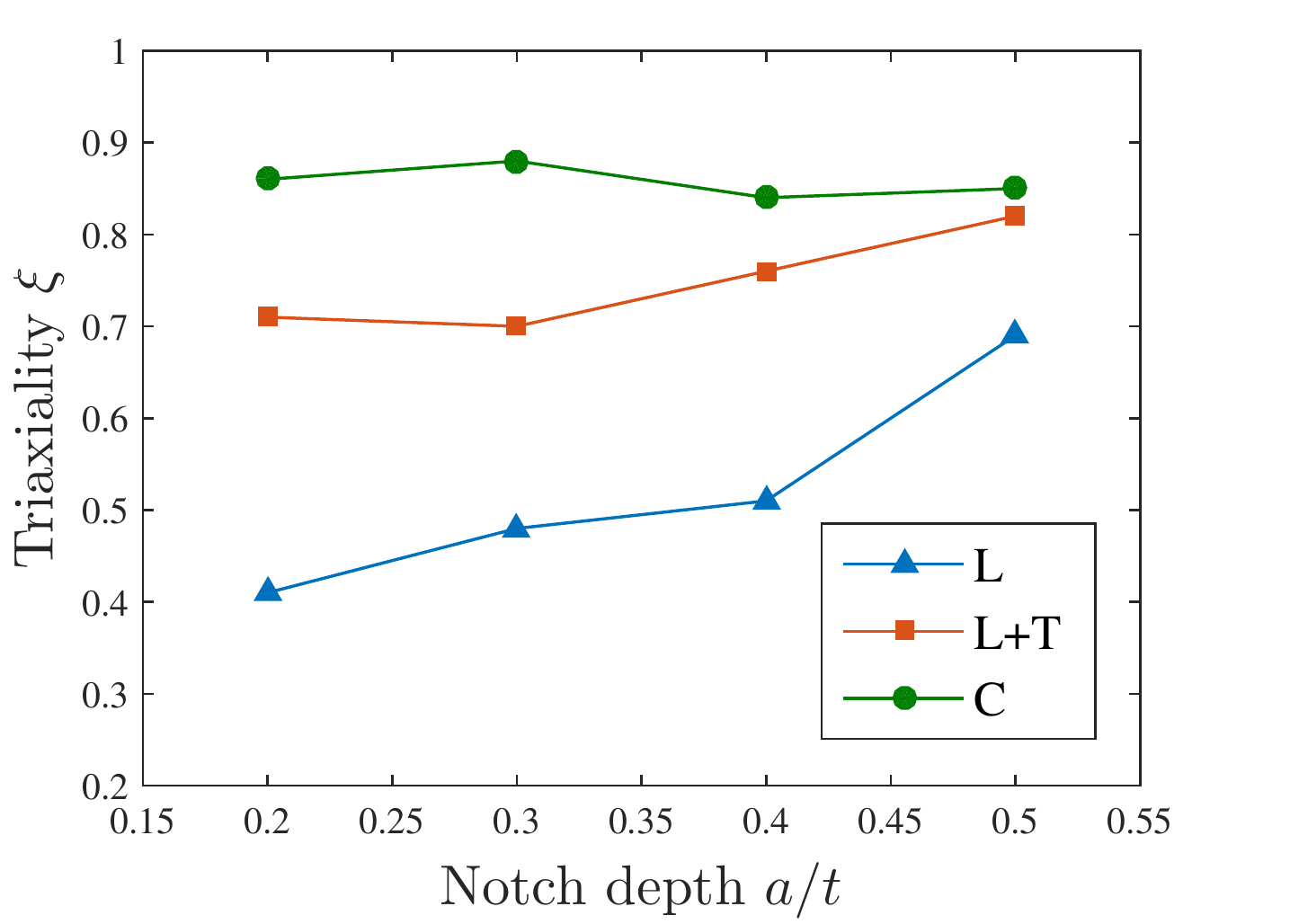}
\caption{Triaxiality levels in the direction of fracture at $r \sigma_y / J =1$ for several notch types, different notch depths and $t=1$ mm.}
\label{fig:Fig1Triax}
\end{figure}

First, the stress triaxiality $\xi$, defined as,
\begin{equation}
\xi=\frac{\sigma_h}{\sigma_e}
\end{equation}

\noindent is computed in the direction of fracture at a normalized distance from the notch tip of $r \sigma_y / J =1$. With $J$ denoting the J-integral, which is computed by means of the domain integral method. Results obtained at the precise instant in which cracking initiates (i.e., $f=f_c$ in all the integration points of an element) are shown in Fig. \ref{fig:Fig1Triax} for the three notch classes considered, different notch depths and a specimen thickness of $t=1$ mm. 

\begin{figure}[H]
\centering
\includegraphics[scale=0.8]{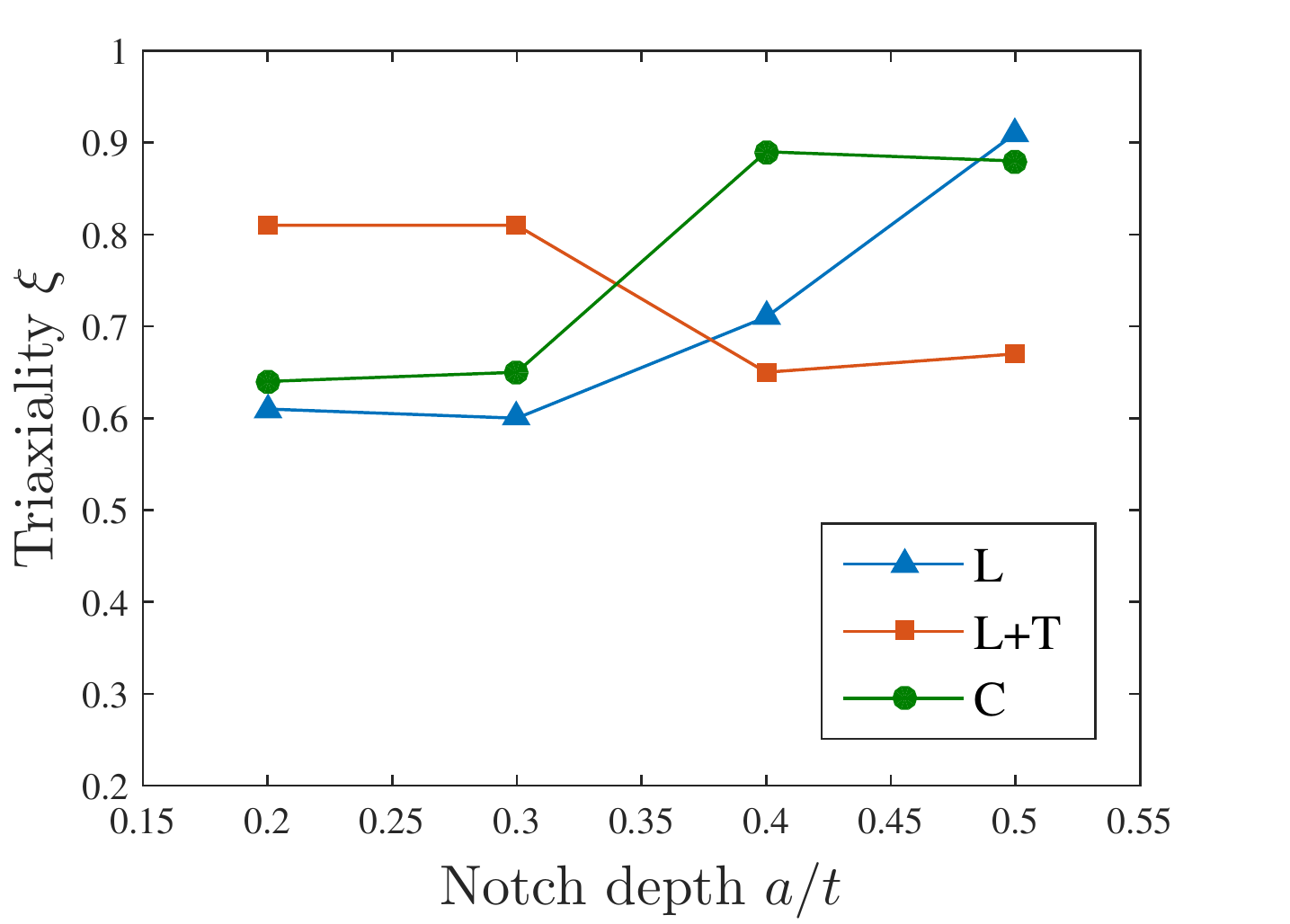}
\caption{Triaxiality levels in the direction of fracture at $r \sigma_y / J =1$ for several notch types, different notch depths and $t=0.5$ mm.}
\label{fig:Fig2Triax}
\end{figure}

Fig. \ref{fig:Fig1Triax} reveals higher stress triaxiality levels in the configurations with a circular notch (C), with the longitudinal notch configuration (L) showing the lowest triaxiality and the longitudinal and transversal notch (L+T) case falling in between. Besides, a high sensitivity to the notch depth is observed in the (L) geometry, while the opposite is shown for the (C) and (L+T) cases. Results are however substantially different when a smaller specimen thickness is assumed ($h=0.5$ mm) as depicted in Fig. \ref{fig:Fig2Triax}.

\begin{figure}[H]
\centering
\includegraphics[scale=0.3]{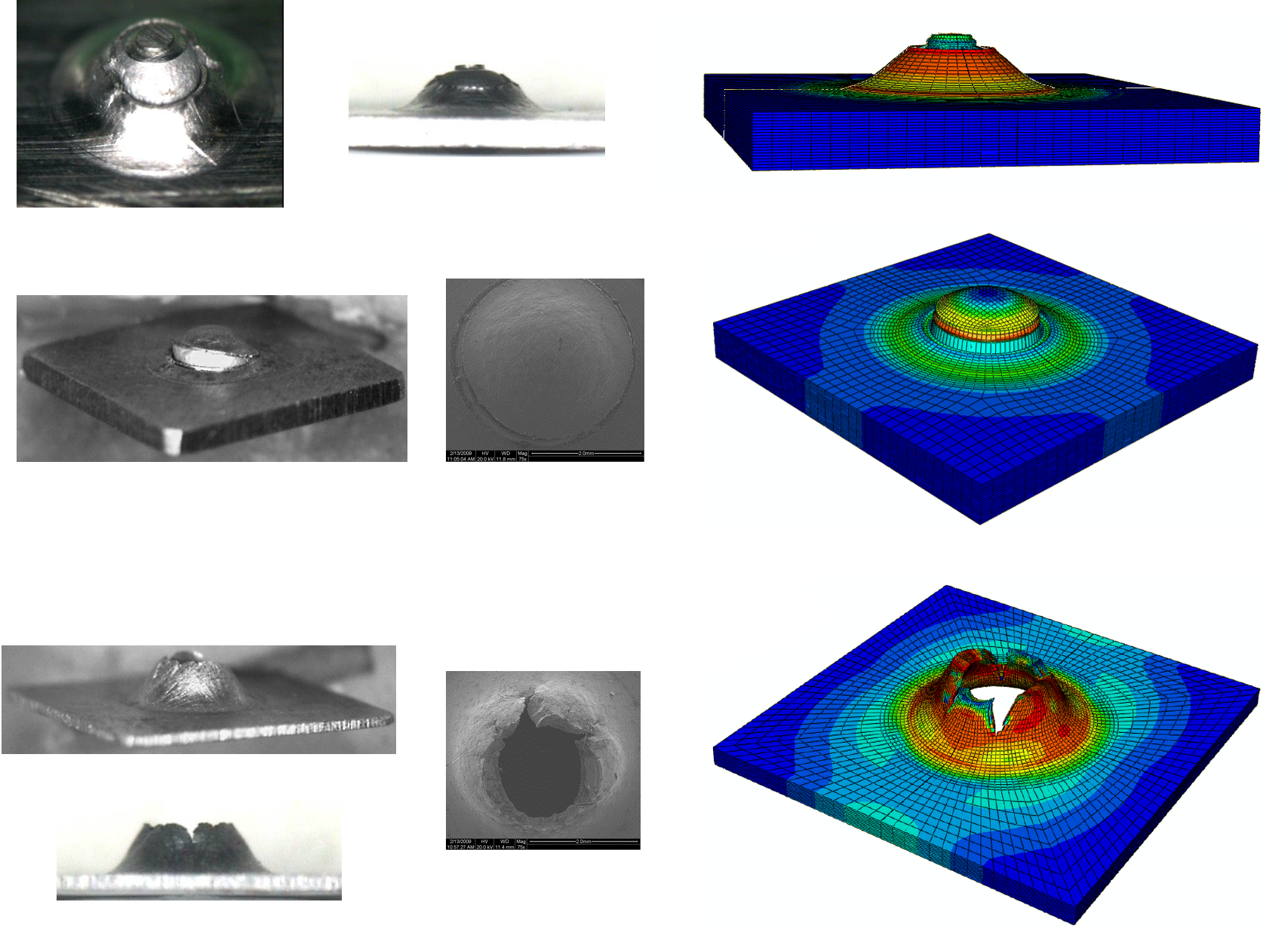}
\caption{Experimental observations and numerical predictions of crack initiation and growth.}
\label{fig:ExptandNum}
\end{figure}

As shown in Fig. \ref{fig:Fig2Triax}, the constraint conditions are now highly dependent on the notch depth, with the longitudinal notch configuration (L) attaining the maximum levels when $a/t=0.5$. An increase in $\xi$ is observed for both (L) and (C) cases when the defect size increases while the opposite trend is shown for the (L+T) configuration. The high sensitivity of the results to the notch depth is explained by the different location of the onset of damage. Thus, in the circular notch configuration, large defect sizes lead to crack initiation sites located at the notch tip, while this is not the case for ratios of $a/t$ lower than 0.4. In all cases the initiation and subsequent propagation of damage trends computed in the numerical model agree with the experimental observations, as depicted in Fig. \ref{fig:ExptandNum}.

\begin{figure}[H]
\centering
\includegraphics[scale=0.8]{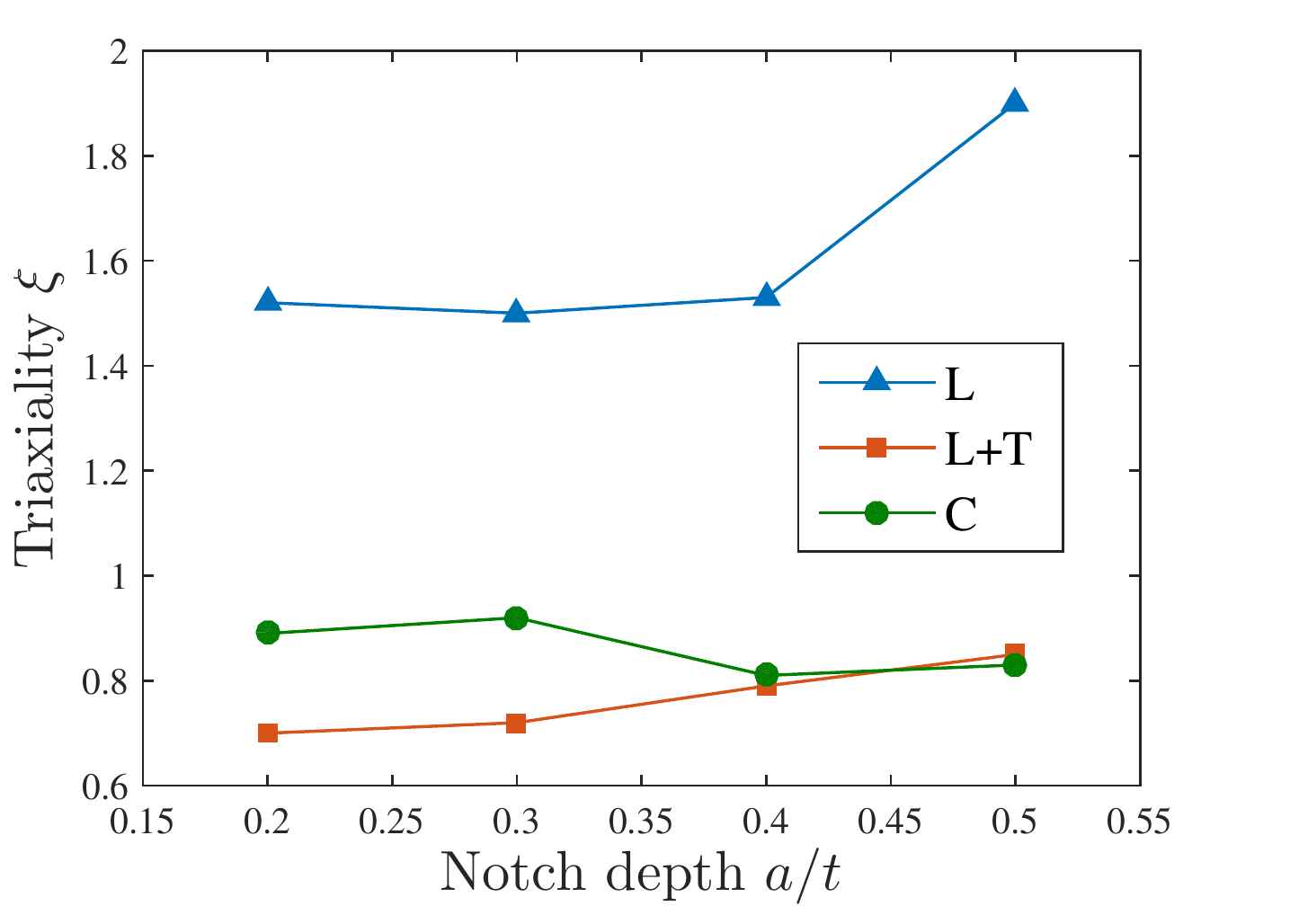}
\caption{Triaxiality levels in the direction of maximum $\xi$ at $r \sigma_y / J =1$ for several notch types, different notch depths and $t=1$ mm.}
\label{fig:Fig3Triax}
\end{figure}

As the location for the onset of damage is highly dependent on the notch to thickness ratio, it may be more appropriate to estimate the triaxiality level in the direction of maximum $\xi$. Fig. \ref{fig:Fig3Triax} shows the results obtained according to this criterion for a thickness of $t=1$ mm and the aforementioned configurations. As in Figs. \ref{fig:Fig1Triax} and \ref{fig:Fig2Triax}, the stress triaxiality is computed at a normalized distance $r \sigma_Y / J =1$ as a function of the ratio between the notch length and the sample thickness.

\begin{figure}[H]
\centering
\includegraphics[scale=0.8]{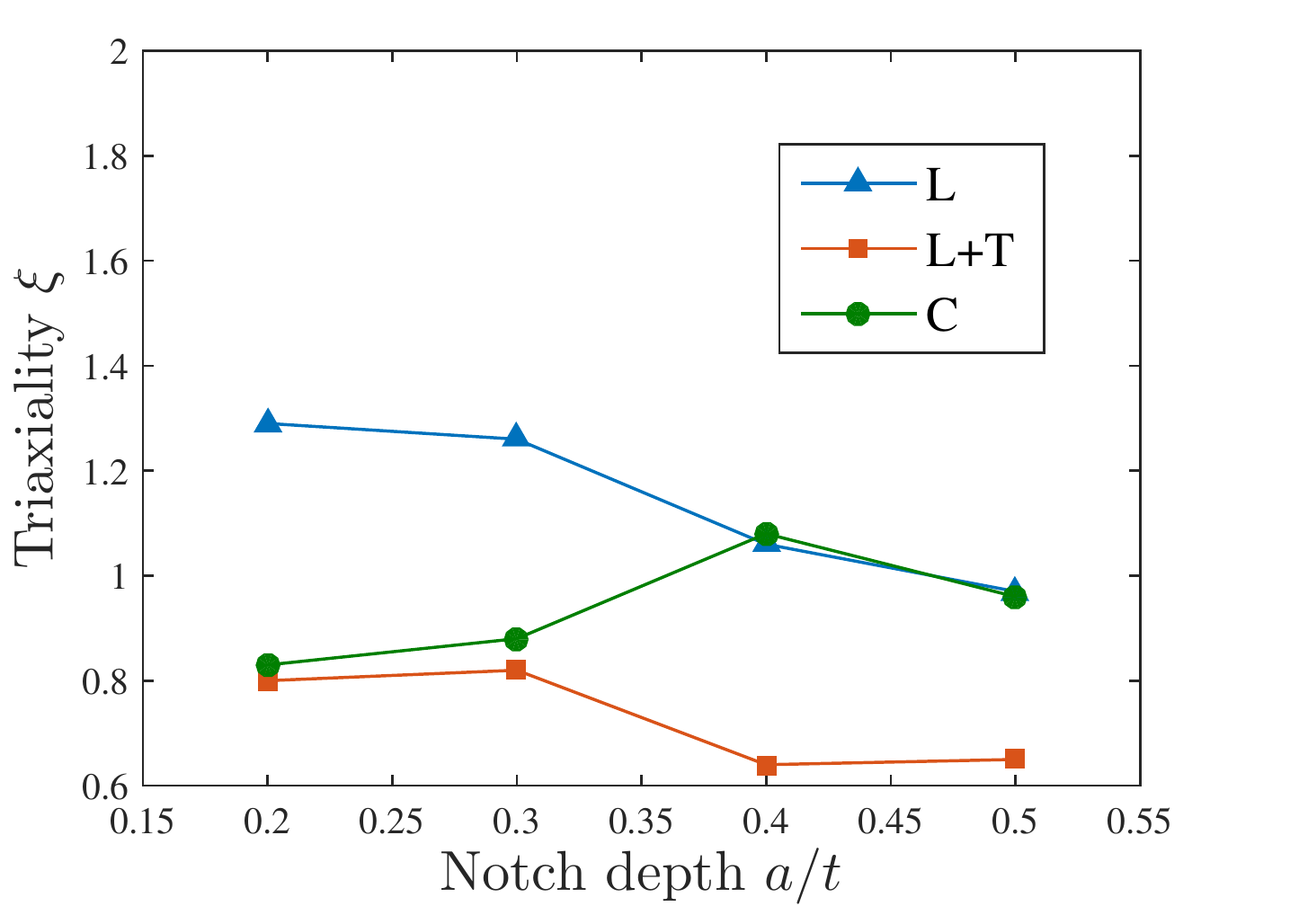}
\caption{Triaxiality levels in the direction of maximum $\xi$ at $r \sigma_y / J =1$ for several notch types, different notch depths and $t=0.5$ mm.}
\label{fig:Fig4Triax}
\end{figure}

Fig. \ref{fig:Fig3Triax} reveals that the triaxiality levels attained with the longitudinal notch configuration (L) are significantly higher than those relevant to the circular (C) and longitudinal and transversal (L+T) notch configurations. A similar trend is observed for a smaller sample thickness, as shown in Fig. \ref{fig:Fig4Triax}. 

Differences between configurations are however smaller when the sample thickness decreases, and the triaxiality levels attained with the longitudinal notch specimen are significantly lower than those shown for $t=1$ mm. Moreover, reducing the thickness of the sample beyond $0.5$ mm could have further implications, as size effects may influence the mechanical response \cite{MB15,MN16}. Highest triaxiality levels seem therefore to be attained with a longitudinal notch for a specimen thickness of $1$ mm.\\

\begin{figure}[H]
\makebox[\linewidth][c]{%
        \begin{subfigure}[b]{0.5\textwidth}
                \centering
                \includegraphics[scale=0.55]{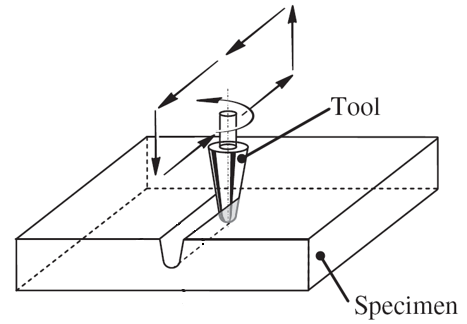}
                \caption{}
                \label{fig:Micromachining}
        \end{subfigure}
        \begin{subfigure}[b]{0.5\textwidth}
                \raggedleft
                \includegraphics[scale=0.51]{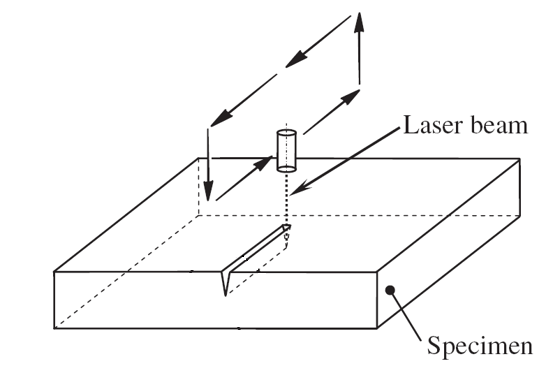}
                \caption{}
                \label{fig:Laser}
        \end{subfigure}
        }
       
        \caption{Schematic view of (a) high-precision micromachining and (b) laser-induced micromachining notch fabrication approaches.}\label{fig:MicromachiningLaser}
\end{figure}

One further aspect to take into consideration is the fabrication process \cite{C11}. Two techniques are mainly being used: (i) high-precision micromachining and (ii) laser-induced micromachining, which will be respectively referred to as micromachining and laser. Each manufacturing procedure leads to a different notch geometry, as shown in Fig. \ref{fig:MicromachiningLaser}. Thus, laser procedures lead to sharper notches with smaller depths than micromachining. Substantial differences are observed in the notch radius as well, with laser-induced techniques leading to values one order of magnitude lower ($e/2=10$ $\mu m$).

\begin{figure}[H]
\centering
\includegraphics[scale=0.8]{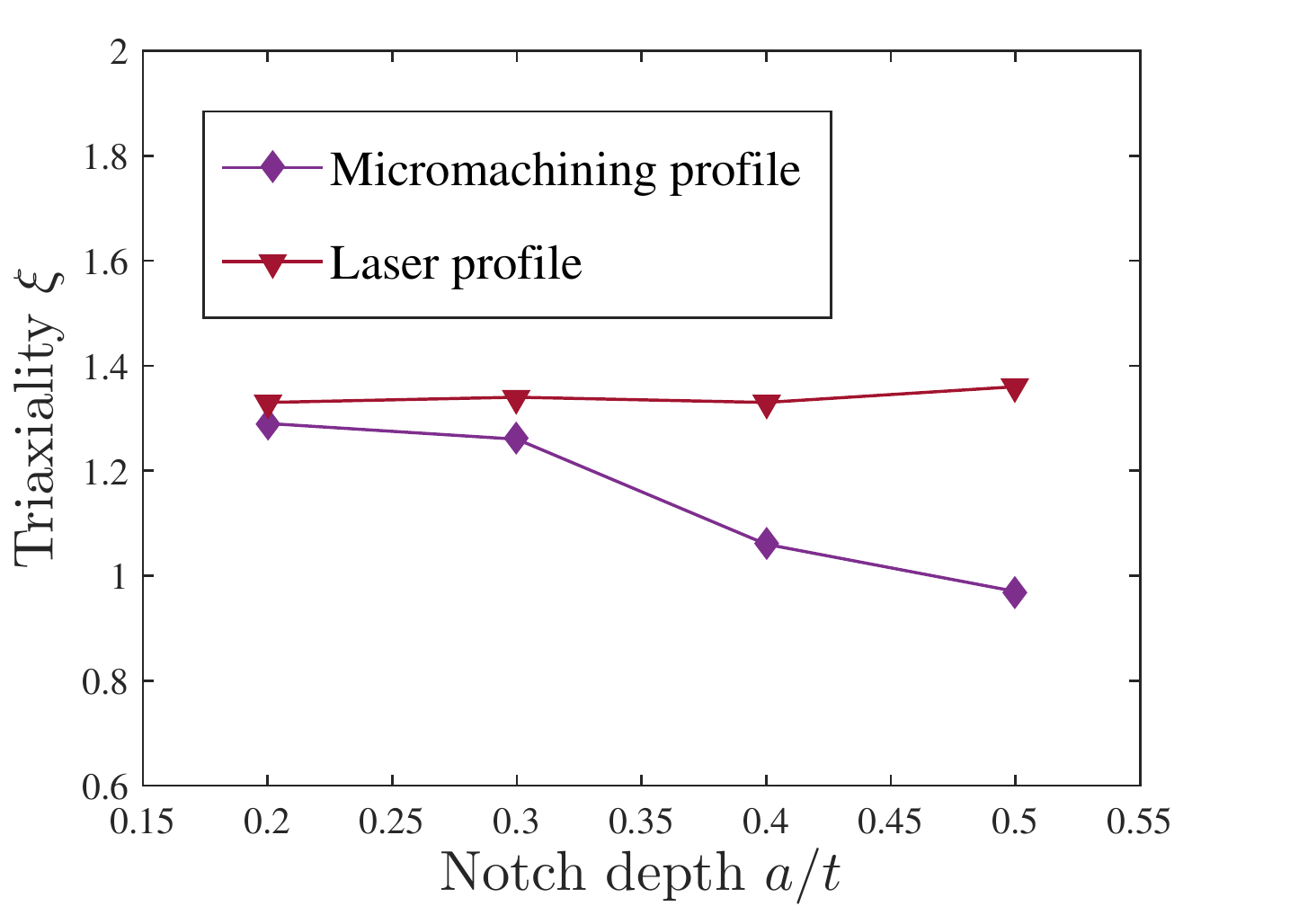}
\caption{Triaxiality levels in the direction of maximum $\xi$ at $r \sigma_y / J =1$ for a longitudinal notch (L) resembling laser and micromachining fabrication techniques, different notch depths and $t=0.5$ mm.}
\label{fig:Fig5Triax}
\end{figure}

The constraint conditions in the direction of maximum triaxiality are examined for notch geometries resembling the outcome of micromachining and laser fabrication approaches and the results are shown in Fig. \ref{fig:Fig5Triax} for the (L) configuration. As shown in the figure, higher triaxialities are obtained with the laser technique, particularly for larger notch depths.  

\begin{figure}[H]
\centering
\includegraphics[scale=0.7]{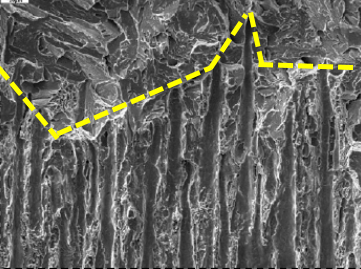}
\caption{Modified SEM image showing the lesser notch uniformity attained with laser-induced micromachining.}
\label{fig:NonuniformLaser}
\end{figure}

However, micromachining leads to a better control of the notching process, which translates in a uniform notch along the specimen length. As shown in Fig. \ref{fig:NonuniformLaser}, this is not the case in laser-based techniques, where less uniformity is observed in the surface finish, with the shape of the notch varying significantly along the specimen length as the depth increases. 

\begin{figure}[H]
\makebox[\linewidth][c]{%
        \begin{subfigure}[b]{0.5\textwidth}
                \centering
                \includegraphics[scale=0.7]{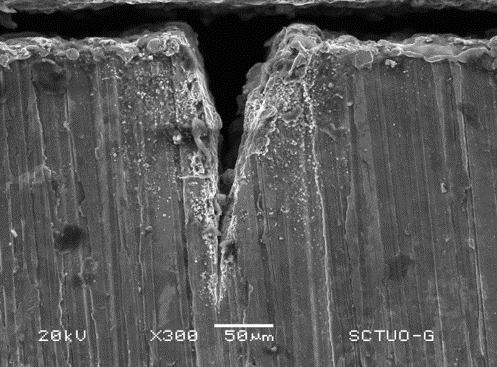}
                \caption{}
                \label{fig:LaserComparative}
        \end{subfigure}
        \begin{subfigure}[b]{0.5\textwidth}
                \raggedleft
                \includegraphics[scale=0.7]{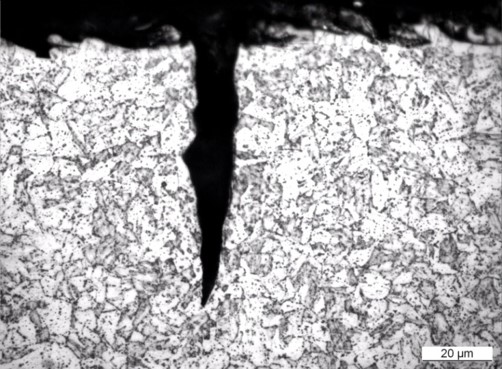}
                \caption{}
                \label{fig:Femtolaser}
        \end{subfigure}
        }
       
        \caption{Cross section of the notch obtained from (a) laser-induced micromachining and (b) femtolaser-induced micromachining.}\label{fig:LaserFemtolaser}
\end{figure}

The aforementioned drawbacks may be alleviated by the use of femtolaser, which allows for a good surface finish and a greater depth accuracy (see Fig. \ref{fig:LaserFemtolaser}). However, the notch losses uniformity far from the center region. Moreover, the manufacturing costs of notched specimens by micromachining are substantially lower than those necessary to introduce defects by means of laser or femtolaser techniques. Consequently, the use of high-precision micromachining is generally recommended.

\subsection{GTN parameters identification through the SPT curve in edge notched specimens}

A novel methodology to extract the parameters that govern the nucleation, growth and coalescence of microvoids in Gurson-Tvergaard-Needleman model is presented. The proposed procedure is employed with SPT specimens partially precracked throughout the thickness and numerical predictions are compared with experimental data for a precipitation hardened martensitic stainless steel of Young's modulus $E=192$ GPa, ultimate strength $\sigma_u=1200$ MPa, yield stress $\sigma_y=1100$ and strain hardening coefficient $n=40$.\\

\begin{figure}[H]
\centering
\includegraphics[scale=0.8]{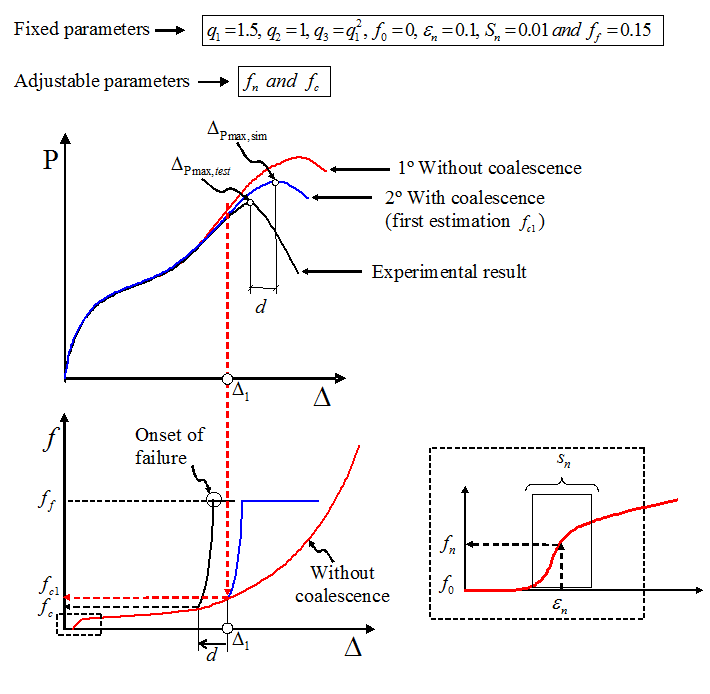}
\caption{Outline of the proposed methodology to identify the GTN parameters from a notched SPT specimen.}
\label{fig:MethodologyUBU}
\end{figure}

The proposed methodology, outlined in Fig. \ref{fig:MethodologyUBU}, aims to assess the critical void volume fraction at the onset of coalescence $f_c$ for given values of the remaining GTN parameters. Thus, following \cite{T81}, $q_1$, $q_2$ and $q_3$ are considered to be respectively equal to $1.5$, $1$ and $2.25$. While, for illustration purposes, it is assumed that $\varepsilon_n=0.1$, $S_n=0.1$ and $f_f=0.15$. The initial void volume fraction $f_0$ is assumed to be equivalent to the volume fraction of intermetallic particles and it is therefore considered to be equal to 0. By having previously fixed the value of $f_n$, which equals 0.01 in the aforementioned case study, the critical void volume fraction $f_c$ can be obtained by means of a number of steps:\\

- Firstly, the nucleation and growth of micro-voids in the SPT is modeled without considering coalescence. In that way, the value of $f_n$ can be easily obtained by fitting the experimental curve.\\

- Afterwards, the punch displacement corresponding to the 90\% of the maximum load in the experimental curve $\Delta_1$ is measured. This quantity is identified as the punch displacement at the onset of failure, as observed in interrupted tests.\\

- The first estimation of the critical void volume fraction $f_{c_1}$ is then obtained from the void volume fraction versus punch displacement curve, as it corresponds to the punch displacement at the onset of failure $\Delta_1$. For this purpose, the void volume fraction variation with punch displacement considered corresponds to the node with higher porosity at the precise instant when the experimental and numerical predictions deviate.\\

- A coalescence-enriched simulation is then performed with the previously extracted value of $f_c$. Afterwards, the difference between the numerical and experimental predictions of the punch displacement at the maximum load level is computed $d=\Delta_{P_{max,sim}}-\Delta_{P_{max,expt}}$.

- Finally, $f_c$ will be estimated from the $f$ versus displacement curve by considering the void volume fraction that corresponds to a punch displacement of $\Delta_1-d$.\\

\begin{figure}[H]
\makebox[\linewidth][c]{%
        \begin{subfigure}[b]{0.6\textwidth}
                \centering
                \includegraphics[scale=0.7]{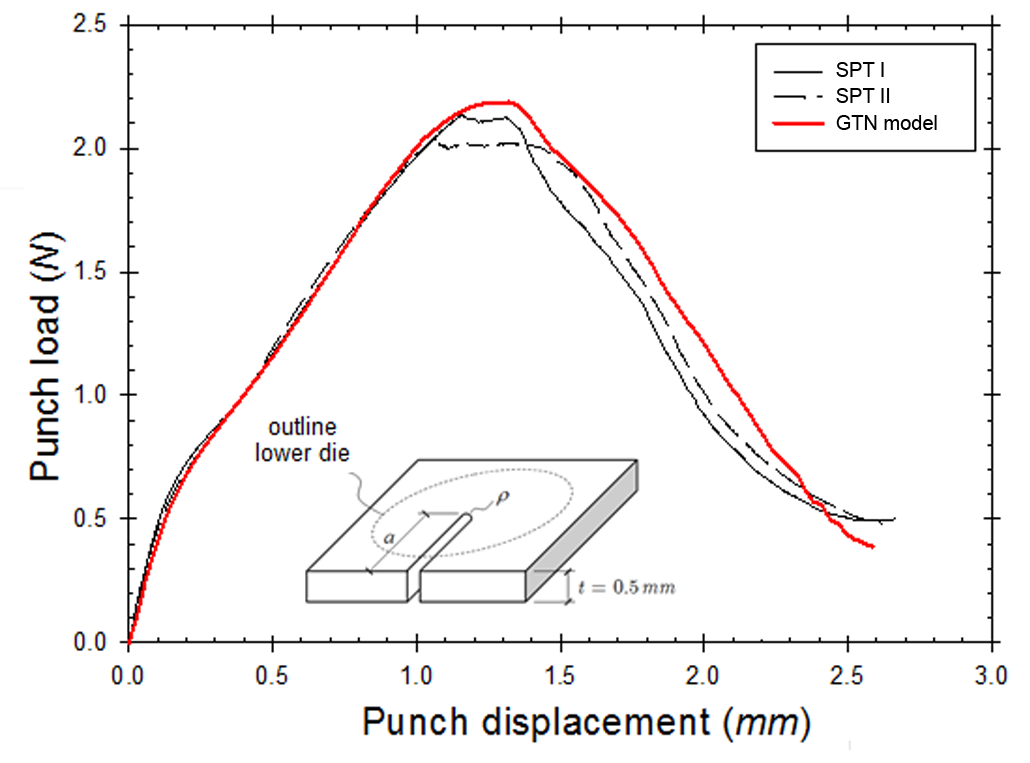}
                \caption{}
                \label{fig:LvsDispCurve}
        \end{subfigure}
        \begin{subfigure}[b]{0.5\textwidth}
                \raggedleft
                \includegraphics[scale=0.12]{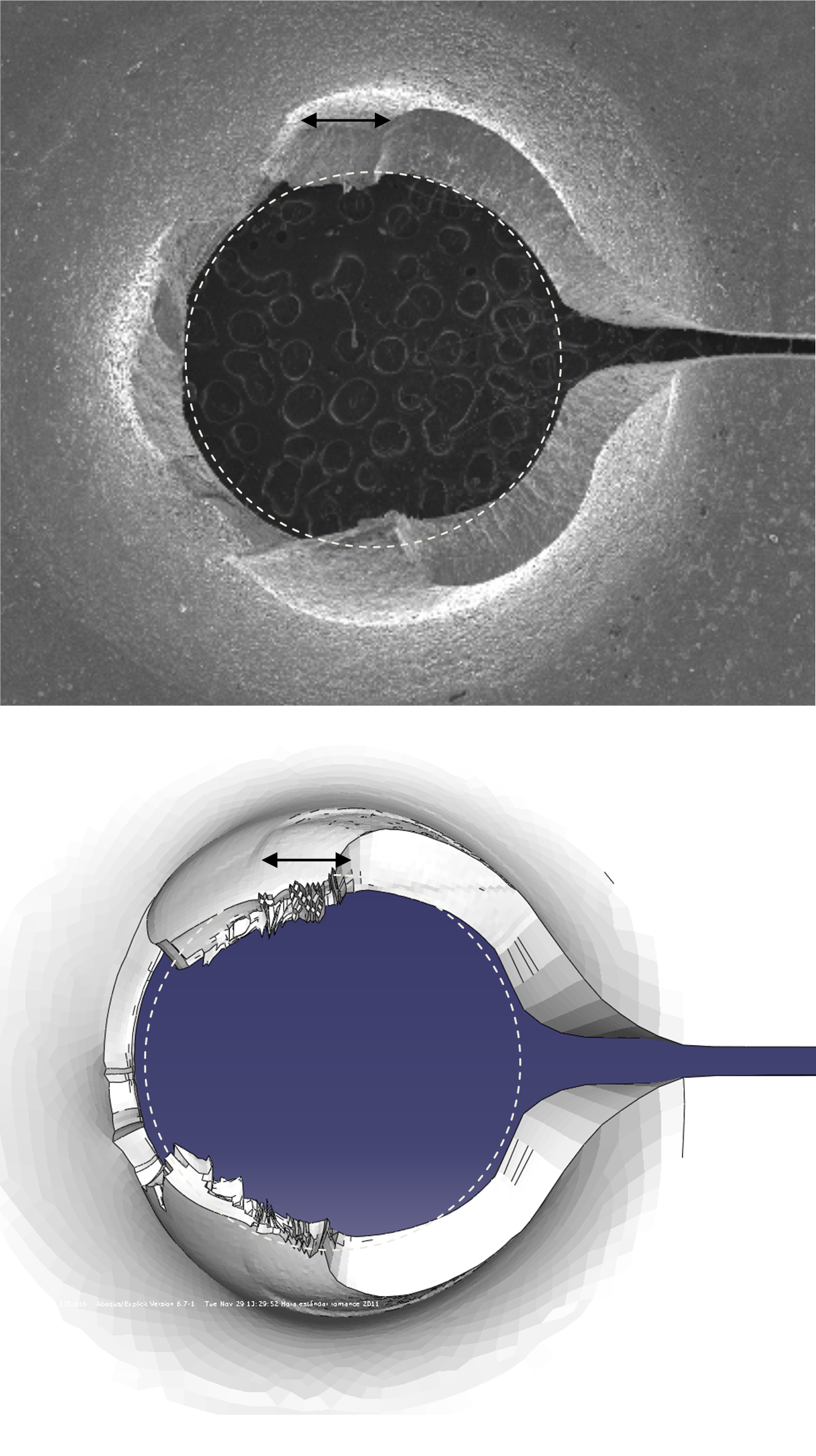}
                \caption{}
                \label{fig:Pictures}
        \end{subfigure}
        }
       
        \caption{Numerical and experimental correlation for SPT expecimens with an edge notch: (a) Load-displacement curve and (b) crack growth predictions. The crack length equals 5 mm.}\label{fig:ResultsUBU}
\end{figure}

The final estimation of $f_c$ allows to accurately capture the experimental trends by means of the GTN model, as shown in Fig. \ref{fig:ResultsUBU}. Two experimental curves are shown (SPT I and SPT II) to give an indication of the experimental scatter.

\subsection{GTN parameters identification through a top-down approach}

While the capabilities of the SPT to accurately estimate mechanical and creep properties are widely known, several uncertainties hinder its use in fracture toughness predictions. Useful insight can be gained by means of micromechanical-based ductile damage models, paving the way for the development of a combined experimental-numerical methodology that will allow to conduct structural integrity evaluations from a very limited amount of material. With this aim, the nucleation and propagation of damage in notched SPT specimens is examined by means of the GTN model. The structural integrity of a CrMoV steel welding joint is assessed by examining the base metal before (CrMoV) and after an intermediate heat treatment of 4 hours at 350$^{\circ}$C (CrMoV IHT). The mechanical properties relevant to both materials are shown in Table \ref{tab:MechProperties}, as extracted from the uniaxial tensile tests. Here, the hardening behavior is fitted with a Hollomon type power law, with $k$ being the strength coefficient and $n$ the strain hardening exponent.

\begin{table}[h]
\caption{Mechanical properties}
\centering
\begin{tabular}{c c c c c c} 
\hline
 & E (GPa) & $\sigma_y$ (MPa) & $\sigma_{u}$ (MPa) & $k$ (MPa) & $n$ \\
 \hline
 CrMoV & 200 & 595 & 711 & 1019 & 0.107 \\
 CrMoV IHT & 210 & 762 & 822 & 1072 & 0.071 \\
 \hline
\end{tabular}
\label{tab:MechProperties}
\end{table}

\begin{figure}[H]
\centering
\includegraphics[scale=0.8]{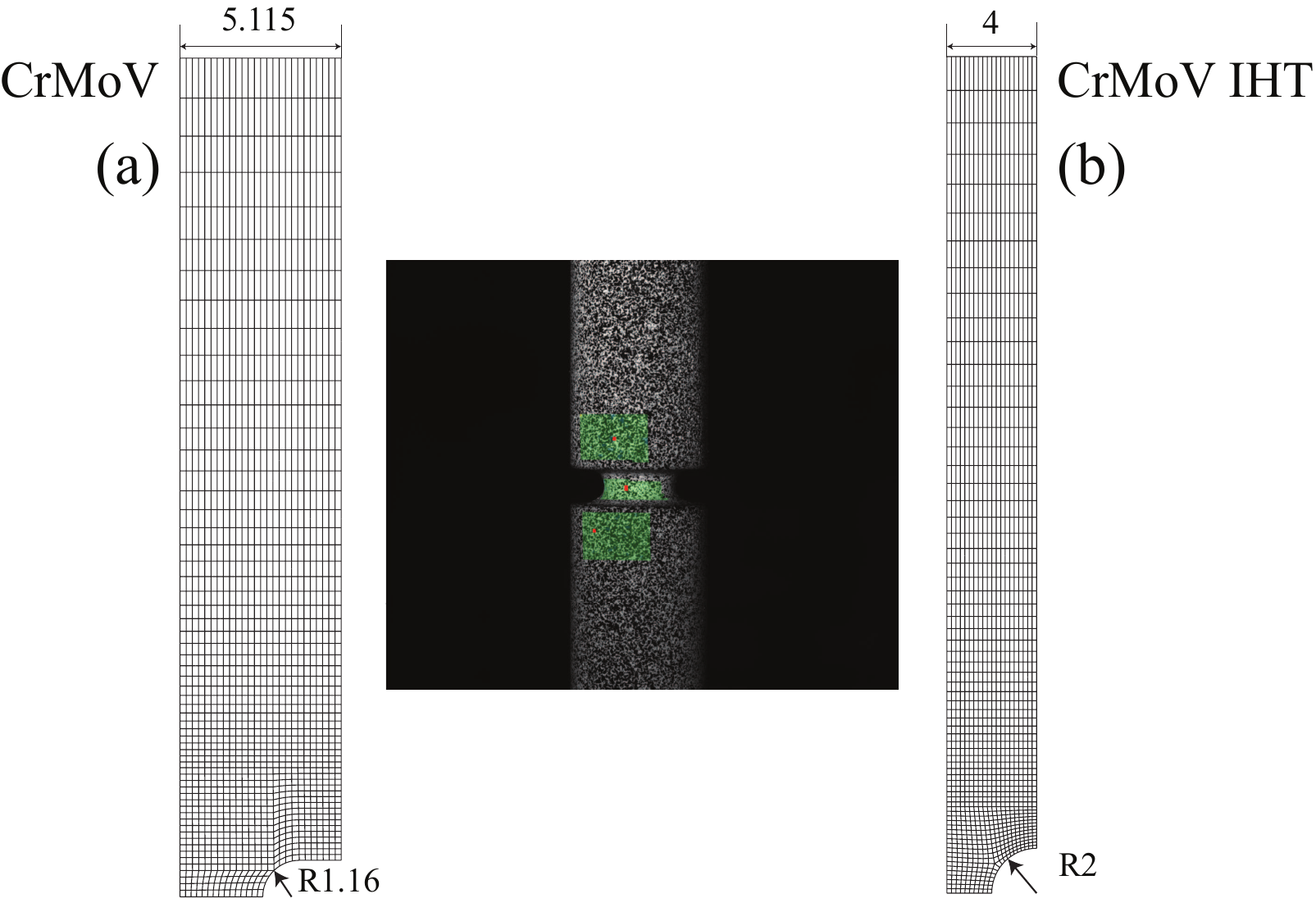}
\caption{Mesh and geometry of the notched uniaxial tensile specimens employed for (a) CrMoV and (b) CrMoV IHT; a representative image of the DIC characterization is also shown. All dimensions are given in mm.}
\label{fig:MeshDIC}
\end{figure}

Following the conclusions extracted from Section \ref{SubSec:TriaxialityStudy}, SPT specimens with a longitudinal notch are employed. The GTN parameters are obtained by fitting through a top-down approach \cite{M16} the load-displacement curve of uniaxial tests in notched round bars. Different specimen geometries are employed in the two material cases considered, being the inner radius of 2.63 mm (CrMoV) and 2 mm (CrMoV IHT). The vertical displacement is accurately measured by means of digital image correlation (DIC), as depicted by the center image of Fig. \ref{fig:MeshDIC}; the samples geometry and the mesh employed are also shown in the figure. Taking advantage of the double symmetry, only one quarter of the specimens is modeled, employing 8-node quadrilateral axisymmetric elements.\\

\begin{figure}[H]
\makebox[\linewidth][c]{%
        \begin{subfigure}[b]{0.6\textwidth}
                \centering
                \includegraphics[scale=0.6]{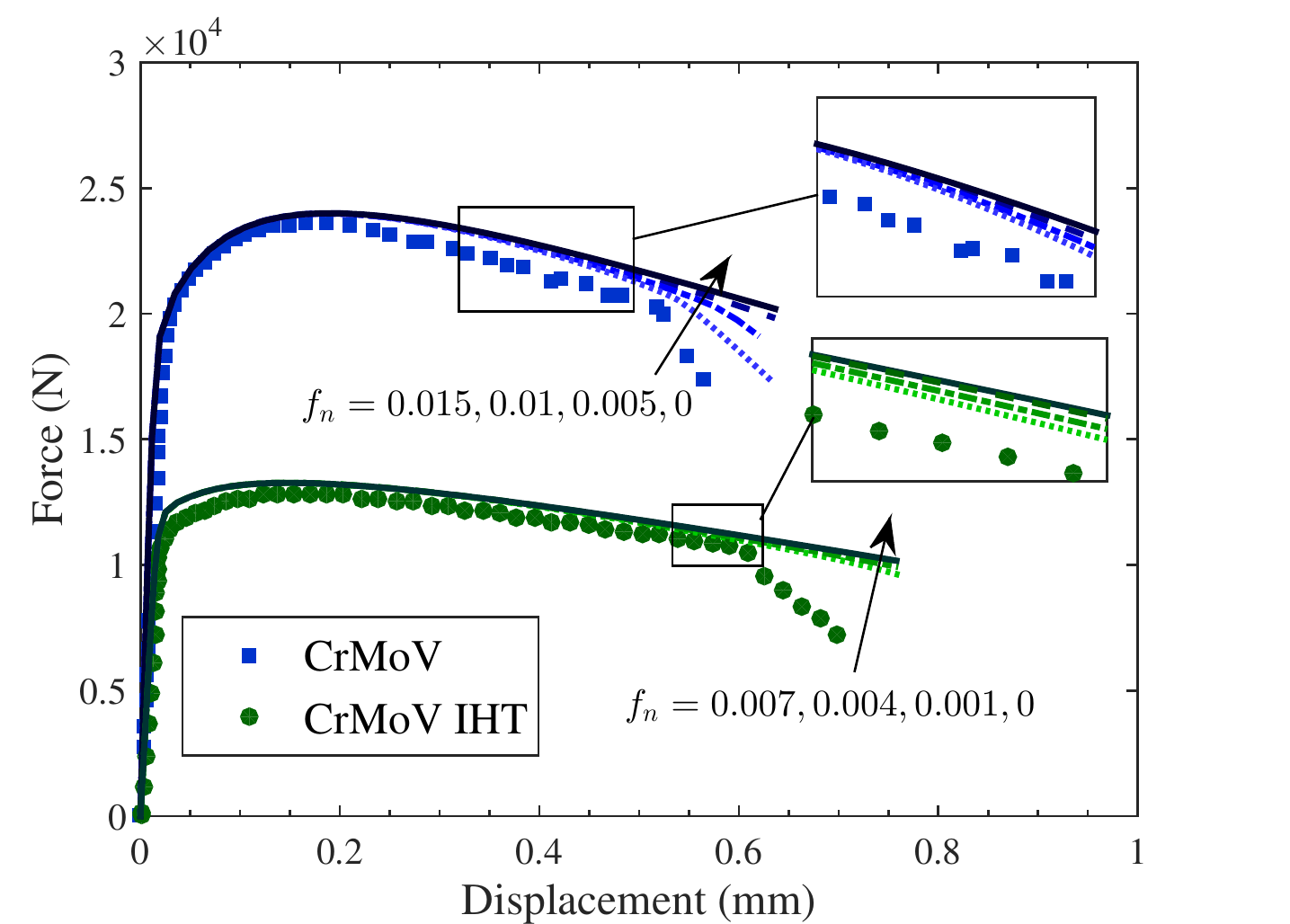}
                \caption{}
                \label{fig:TopDownA}
        \end{subfigure}
        \begin{subfigure}[b]{0.6\textwidth}
                \raggedleft
                \includegraphics[scale=0.6]{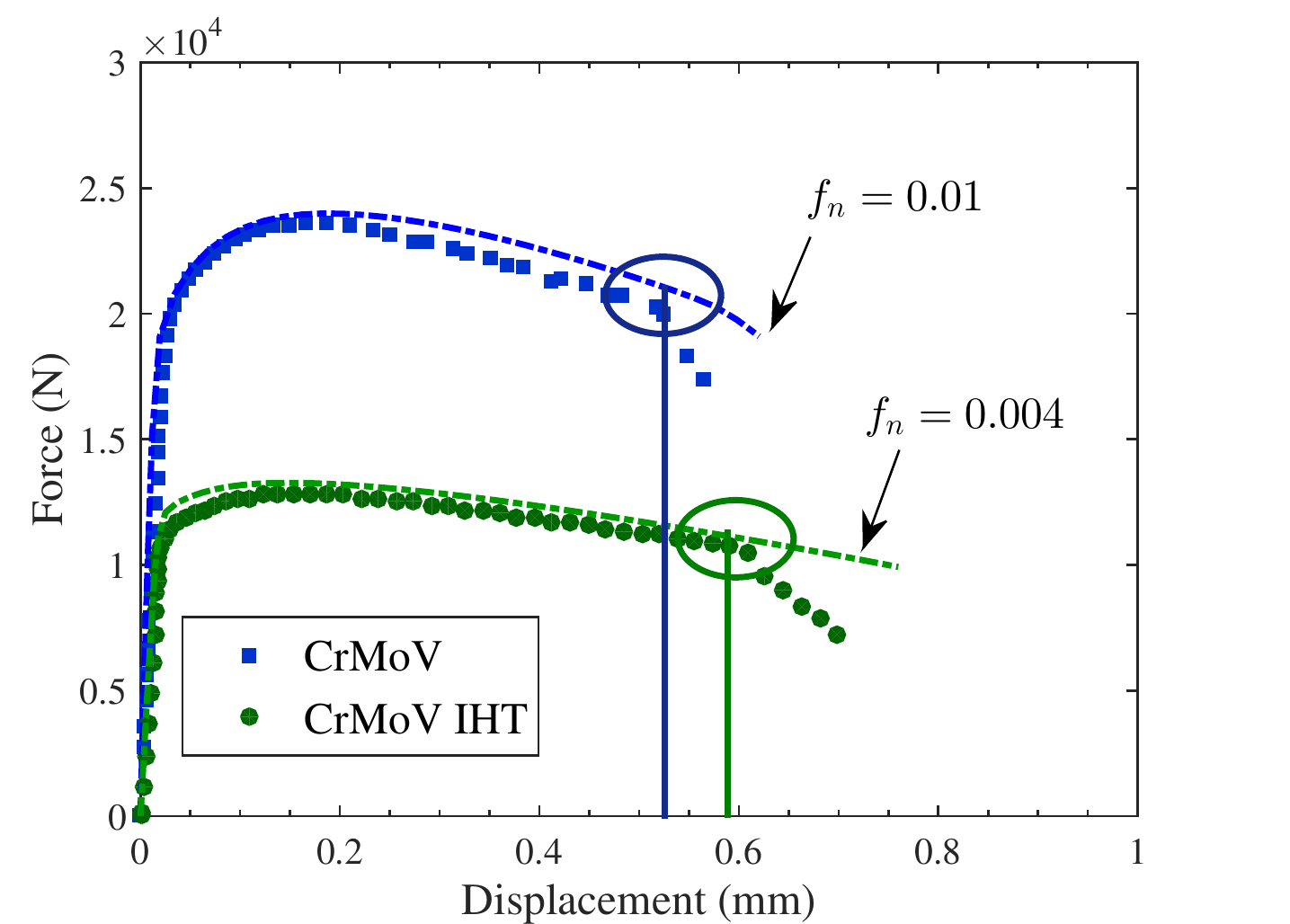}
                \caption{}
                \label{fig:TopDownB}
        \end{subfigure}}

\makebox[\linewidth][c]{%
        \begin{subfigure}[b]{0.6\textwidth}
                \centering
                \includegraphics[scale=0.6]{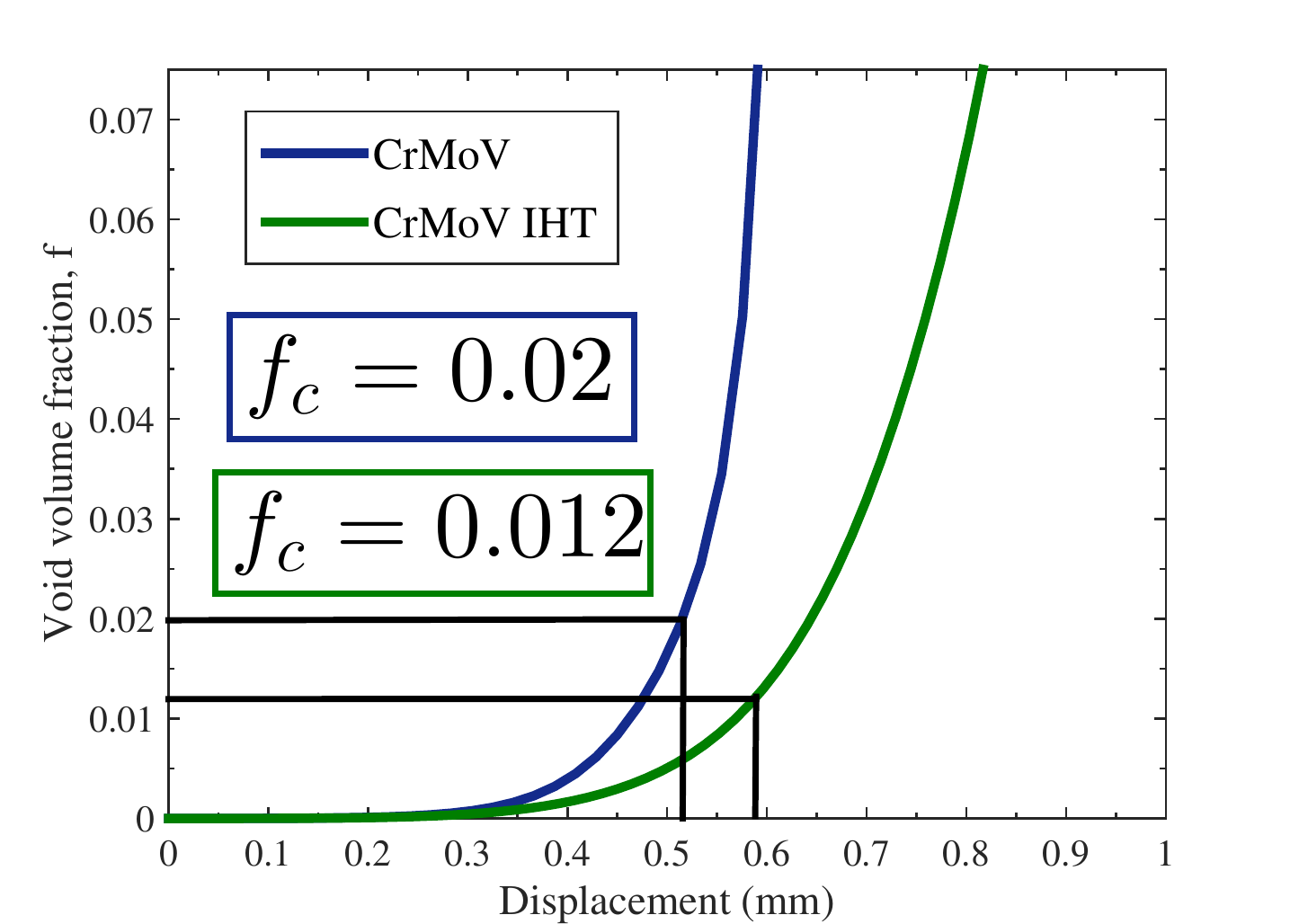}
                \caption{}
                \label{fig:TopDownC}
        \end{subfigure}
        \begin{subfigure}[b]{0.6\textwidth}
                \raggedleft
                \includegraphics[scale=0.6]{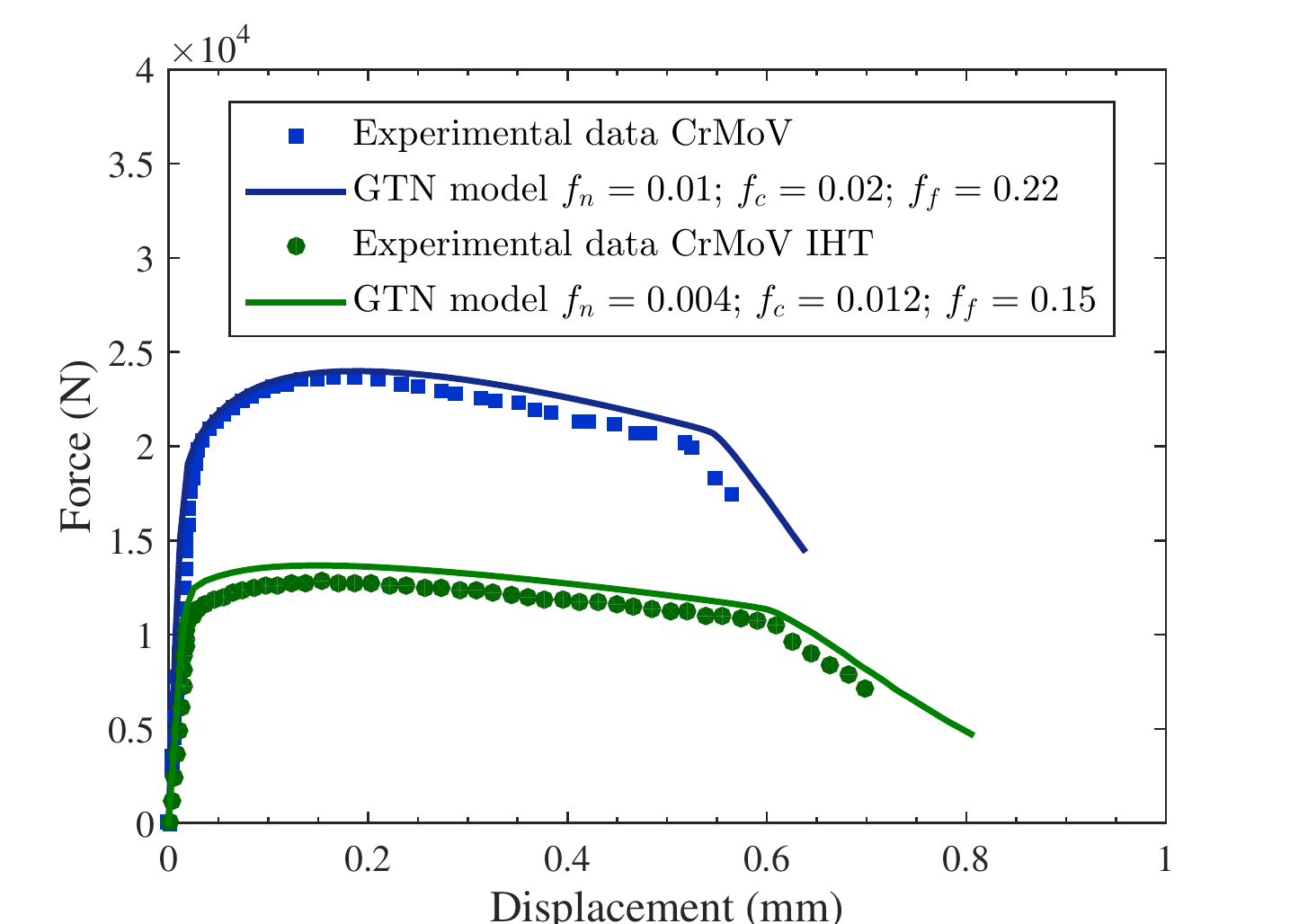}
                \caption{}
                \label{fig:TopDownD}
        \end{subfigure}
        }       
        \caption{Outline of the top-down approach: (a) Experimental data and numerical predictions for different values of $f_n$, (b) identification of the sudden load drop associated with void coalescence, (c) void volume fraction in the center of the specimen versus displacement for the chosen value of $f_n$, (d) numerical damage simulation.}\label{fig:TopDown}
\end{figure}

GTN parameters are obtained by first assuming $q_1=1.5$, $q_2=1.0$, $q_3=2.25$ \cite{T81} and $\varepsilon_n=0.3$, $S_n=0.1$ \cite{CN80}; while $f_0$, $f_n$, $f_c$ and $f_f$ are identified by calibrating with experiments through a top-down approach. As in the previous section, a zero initial void volume fraction $f_0=0$ is adopted, as it is assumed to correspond to the volume fraction of intermetallic particles. The remaining parameters ($f_n$, $f_c$ and $f_f$) are identified from the experimental load-displacement curve of the notched uniaxial samples, as outlined in Fig. \ref{fig:TopDown}. First, the void volume fraction of nucleating particles $f_n$ is obtained by correlating the experimental data with the numerical results obtained without considering void coalescence. Afterwards (Figs. \ref{fig:TopDownB} and \ref{fig:TopDownC}), the critical void volume fraction $f_c$ is identified by assuming that it corresponds with the rapid loss in strength characteristic of void coalescence. And lastly, the slope of the experimental curve after the onset of failure determines the value of $f_f$ (Fig. \ref{fig:TopDownD}).

Damage parameters obtained for the base metal before and after the intermediate heat treatment are displayed in Table \ref{tab:GursonParameters}. By employing uniaxial tensile tests on notched specimens for the GTN parameter identification it is possible to clearly establish the location of the onset of damage and accurately measure the displacement through the DIC technique. 

\begin{table}[H]
\caption{Ductile damage modeling parameters (GTN model) obtained from a notched tensile test through a top-down approach}
\centering
\begin{tabular}{c c c c c c c c c c} 
\hline
 & $q_1$ & $q_2$ & $q_3$ & $f_0$ & $\varepsilon_n$ & $S_n$ & $f_n$ & $f_c$ & $f_f$ \\
 \hline
 CrMoV & 1.5 & 1.0 & 2.25 & 0 & 0.3 & 0.1 & 0.01 & 0.02 & 0.22 \\
 CrMoV IHT & 1.5 & 1.0 & 2.25 & 0 & 0.3 & 0.1 & 0.004 & 0.012 & 0.15 \\
 \hline
\end{tabular}
\label{tab:GursonParameters}
\end{table}

\begin{figure}[H]
\makebox[\linewidth][c]{%
        \begin{subfigure}[b]{0.55\textwidth}
                \centering
                \includegraphics[scale=0.55]{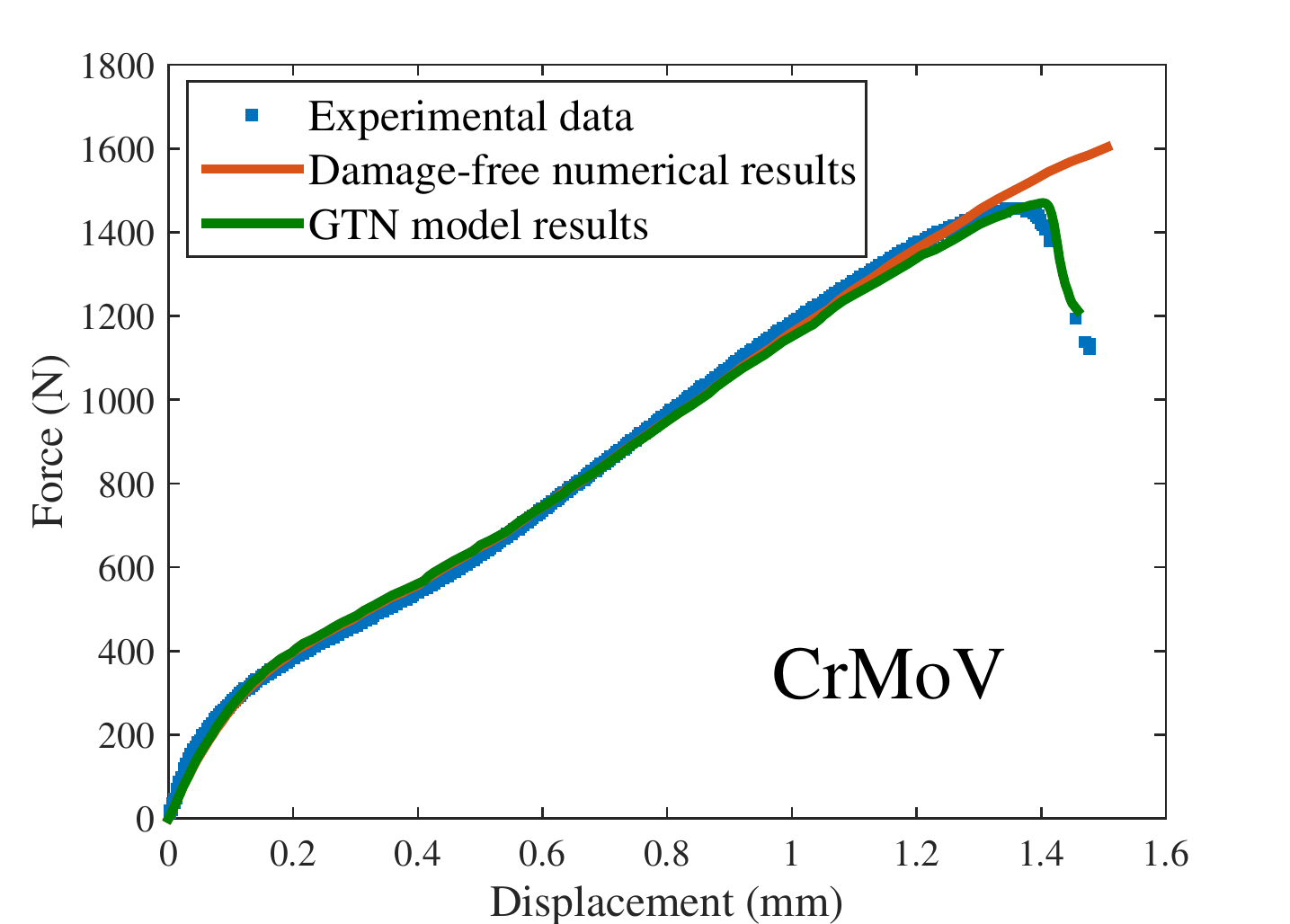}
                \caption{}
                \label{fig:LoadvsDispCrMoV}
        \end{subfigure}
        \begin{subfigure}[b]{0.55\textwidth}
                \raggedleft
                \includegraphics[scale=0.55]{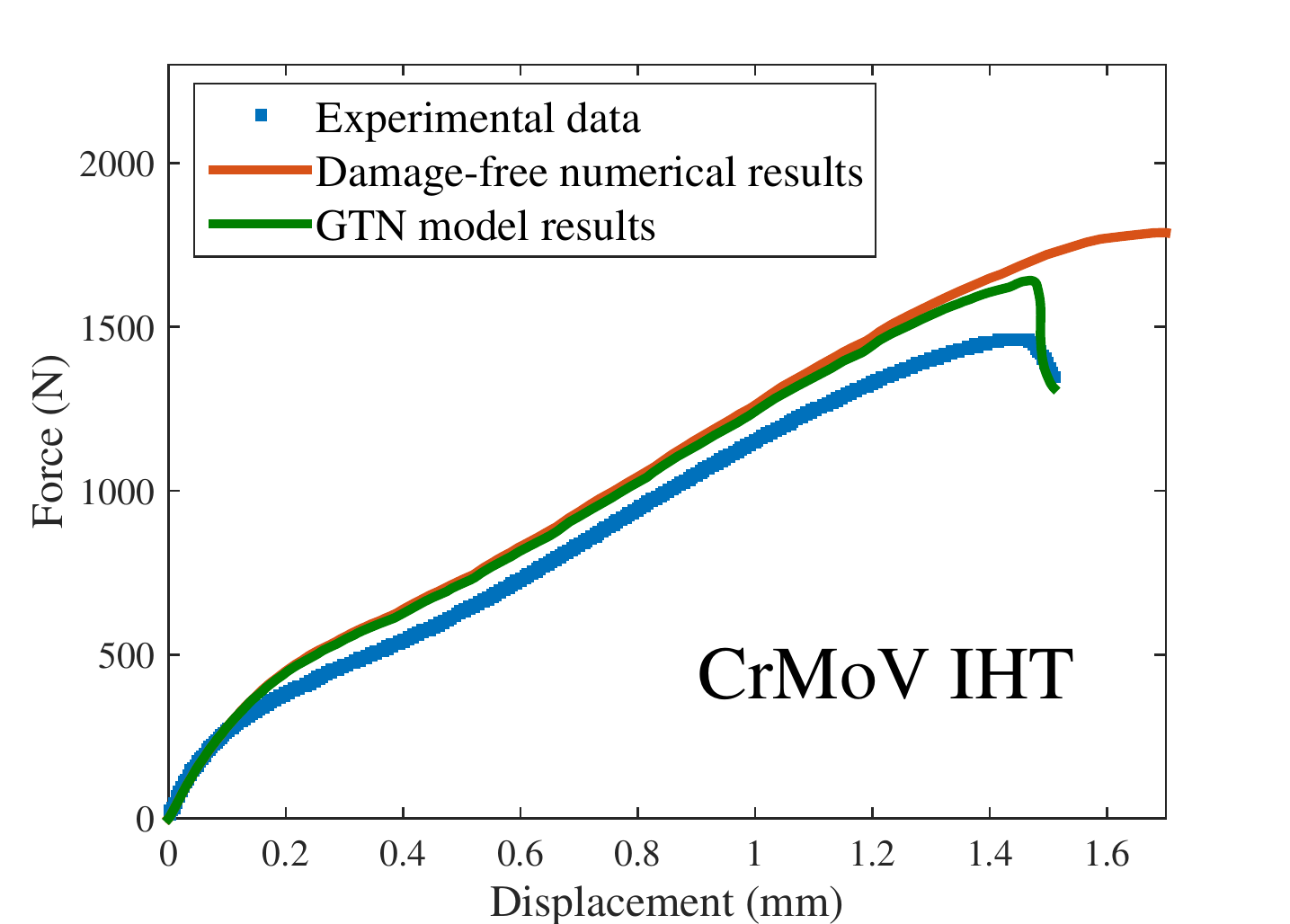}
                \caption{}
                \label{fig:LoadvsDispCrMoV1}
        \end{subfigure}
        }
       
        \caption{SPT experimental and numerical (with and without damage) load-displacement curves for (a) CrMoV and (b) CrMoV IHT}\label{fig:NumResults}
\end{figure}

\begin{figure}[H]
\centering
\includegraphics[scale=0.3]{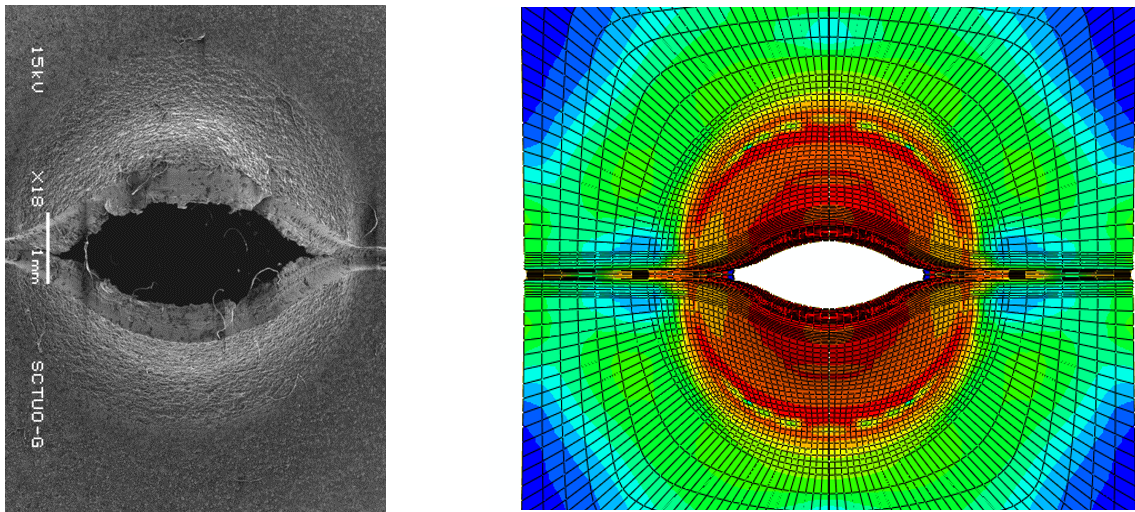}
\caption{Different notched SPT specimens examined}
\label{fig:NumResults1}
\end{figure}

The GTN model parameters shown in Table \ref{tab:GursonParameters} are subsequently employed to model nucleation, growth and coalescence in the SPT. The experimental and numerical results obtained for both materials are shown in Figs. \ref{fig:NumResults} and \ref{fig:NumResults1}. Fig. \ref{fig:NumResults} shows the damage-enhanced numerical predictions along with the experimental data and the conventional elasto-plastic simulations; GTN results precisely follow the experimental curve in both cases, showing the good performance of the top-down methodology employed.

In Fig. \ref{fig:NumResults1} one can easily observe that the onset of damage and subsequent propagation is accurately captured by the numerical model. This is particularly useful for the development of new methodologies for fracture toughness assessment within the SPT, as it allows to identify crack propagation patterns and measure the crack tip opening displacement \cite{M16}.

\section{Conclusions}
\label{Concluding remarks}

Ductile damage modeling within notched SPT specimens has been thoroughly examined. The different perspectives adopted have been reviewed and the choice of an appropriate notch geometry has been extensively studied, from both triaxiality and manufacturing considerations.\\

Particular emphasis is placed on the identification of the GTN model parameters. On the one hand, a novel methodology is proposed with the aim of enabling ductile damage modeling from the load versus punch displacement curve. On the other hand, a top-down approach is employed to gain insight into the mechanisms of crack growth in the SPT, with the ultimate goal of developing an standardized procedure to accurately assess fracture toughness from small scale experiments.

\section{Acknowledgments}
\label{Acknowledge of funding}

The authors gratefully acknowledge financial support from the Ministry of Economy and Competitiveness of Spain through grant MAT2014-58738-C3.




\end{document}